# Reflections on Shannon Information:

## In search of a natural information-entropy for images

Kieran G. Larkin

Nontrivialzeros Research, 22 Mitchell Street, Putney, NSW 2112,

Australia

iresearch@nontrivialzeros.net

2 September 2016

## Abstract

It is not obvious how to extend Shannon's original information entropy to higher dimensions, and many different approaches have been tried. We replace the English text symbol sequence originally used to illustrate the theory by a discrete, bandlimited signal. Using Shannon's later theory of sampling we derive a new and symmetric version of the second order entropy in 1D. The new theory then naturally extends to 2D and higher dimensions, where by *naturally* we mean simple, symmetric, isotropic and parsimonious. Simplicity arises from the direct application of Shannon's joint entropy equalities and inequalities to the gradient (del) vector field image embodying the second order relations of the scalar image. Parsimony is guaranteed by halving of the vector data rate using Papoulis' generalized sampling expansion. The new 2D entropy measure, which we dub *delentropy*, is underpinned by a computable probability density function we call *deldensity*. The deldensity captures the underlying spatial image structure and pixel co-occurrence. It achieves this because each scalar image pixel value is nonlocally related to the entire gradient vector field. Both deldensity and delentropy are highly tractable and yield many interesting connections and useful inequalities. The new measure explicitly defines a realizable encoding algorithm and a corresponding reconstruction. Initial tests show that delentropy compares favourably with the conventional intensity-based histogram entropy and the compressed data rates of lossless image encoders (GIF, PNG, WEBP, JP2K-LS and JPG-LS) for a selection of images. The symmetric approach may have applications to higher dimensions and problems concerning image complexity measures.

# Structure of Paper

(just for layout purposes; to be deleted before final version)

1. Introduction and Motivation
2. Related Work: A Partial History of Image Entropy
3. Re-interpreting Shannon's 1D symbol entropy
4. Extending Shannon's re-interpreted entropy into 2D
5. Some properties of the 2D deldensity function
6. Implementation details
7. Examples and comparisons
8. Discussion and Future Research

   Acknowledgements

   References

   Appendices

   A1:   Image reconstruction from partial derivatives sampled at half-rate

   A2:   Prior Incarnations of The Gradient Mapping or Deldensity

   A3:   Entropy of vector and phase images in 2D

# 1    Introduction and Motivation

## Context

It is almost 70 years since Claude Shannon [1,2] introduced the mathematically precise measure of the amount of information that is exchanged in the process of communication. Shannon found that his information measure had exactly the same formulation as a concept familiar to physicists working on thermodynamics, namely entropy. Shannon's information ideas are now commonplace in our digital world and explicitly used in the design of communication, compression and cryptographic systems

## Need

Shannon's original papers (and the popular exposition [3]) are remarkably wide-ranging, covering zeroth, 1st, 2nd, and 3rd order theory for encoding (English) text, as well as 1st and 2nd order theory for encoding (English) words. Shannon goes on to consider joint entropies and even multidimensional entropies based on multidimensional probability density functions, as well as continuous and discrete bandlimited signals. However, Shannon stopped short of computing the entropy for specific bandlimited signals with correlations, except for the generic case of filtered white noise. Also missing is any reference to 2D signals with structure, such as images.

Subsequently many authors have proposed ingenious and variously effective ways to compute "entropy" or complexity for images or other two dimensional objects such as lattices or geographical patterns (see next section on prior art). Some of the early and simple definitions failed to correspond in any way to humanly perceived complexity. And then some of the more sophisticated measures seem to fail simple tests of symmetry or invariance. Researchers have found that it is very hard to take Shannon's original, essentially 1D, theory and find a way to extend it naturally into higher dimensions.

There is also the question of why anyone should be interested in a good definition of information-entropy for images. Surely a good definition of the amount information that can be communicated (like Shannon's) must necessarily define an encoding and decoding scheme that approaches the minimum redundancy afforded by an entropy measure? Many current 2D entropy measures bear little relation to other measures of the intrinsic information content such as lossless compressed file size.

## Our objective

The goal of a "natural" realization of entropy for 2D (and more specifically digital images in our case) may have remained out of reach because of seemingly intractable symmetry problems. These problems need to be examined in detail. We have identified the main assumptions in Shannon's original theory, then separated them into essential and incidental assumptions. The non-essential assumptions have been found to contain a number of mechanisms that prevent a fully symmetric application even in 1D bandlimited signals. A natural extension to higher dimensions cannot proceed until the 1D case has been reinterpreted. Note that it is unclear whether Shannon himself interpreted his theory in this way. In a sense it is the choice of data type used in his examples (English text) that obscures its re-application to bandlimited signals and images.

The key factor is that the second order estimate of information entropy is generally presented in a causal (asymmetric) form, whereas the zeroth and first order estimates are simply blind to data ordering or sequence. The asymmetry prevents attempts to extend the second order method to higher dimensions in much the same way that the 1D Hilbert transform prevented the definition of a 2D analytic signal before 2001 [4,5]. We derive a symmetric second order information measure in 1D and show how it naturally extends to 2D and higher dimensions. Some interesting new problems arise in 2D, but they turn out to be solvable by an extension [6] of Shannon's later contribution: sampling theory [7].

The new theory naturally satisfies the requirements of symmetry, invariance, parsimony, and simplicity and correlates well with human perceptions of complexity in signals and images.

## Preview

The paper starts with a partial review of research in the areas of signal and image entropy measures.

We review Shannon's original assumptions and suggest a symmetrical re-interpretation of second order symbol entropy. The remainder of the paper introduces the new definitions and concepts, first in 1D, then 2D. The technique is applied to a series of images and compared to other entropy measures and the best lossless compressors. We then discuss connections with other areas of current research and future opportunities. We have added an appendix outlining non-redundant sampling theory for vector field images. A second appendix presents further properties and connections of the new 2D density distribution. A third appendix outlines important connections with information measures for phase and vector images.

Note that this paper introduces a potentially controversial approach to 2D information-entropy. It remains essentially a proof of concept and we hope and expect future research to build, prove, challenge, innovate and generally develop the basic (but very promising) techniques outlined here.

## 2   Related Work: A Partial History of Image Entropy

When Shannon [1] introduced the concept of the minimum amount of information (in bits per symbol) needed to encode and communicate a text message he inextricably intertwined the concept with thermodynamic entropy.  It seems the infamous comment of von Neumann [8,9] has been vindicated:

> *You should call it entropy, for two reasons. In the first place your uncertainty function has been used in statistical mechanics under that name. In the second place, and more importantly, no one knows what entropy really is, so in a debate you will always have the advantage.*

Shannon first acknowledged Tukey's seminal suggestion for the word *bit* to denote a base-2 logarithmic unit of information.  Then, in his Theorem 2 Shannon derived an equation for the total information in a message and observed that it had the same form as Boltzmann's statistical mechanics definition of entropy.  However, the concept of entropy is problematic , even when restricted to thermodynamics..  For example Jaynes [10] observed in 1965:

> *It is clearly meaningless to ask "What is the entropy of the crystal?" unless we first specify the set of parameters which define its thermodynamic state.*

Furthermore Jaynes quotes E.P. Wigner:

> *Entropy is an anthropomorphic concept.*

Subsequently extending the idea of Shannon's information entropy to signals and images has been fraught with the dangers of applying analogies to disparate concepts.  Bracewell [11] noted enigmatically:

> *There is a certain air of mystery surrounding the definition of image entropy.*

Hamming [12] summarised it perfectly:

> *The name "entropy" is used because the same mathematical form arises in thermodynamics and in statistical mechanics, and hence the word "entropy" gives an aura of importance which is not justified in the long run. The same mathematical form does not imply the same interpretation of the symbols!*

Shannon originally worked with symbols (text characters or letters at the zeroth order approximation) and then small groups of symbols or words.  But Shannon's symbols are not directly comparable to (discrete) signal values.  This is because symbols have no natural ordering (the alphabet is essentially an arbitrary ordering, as is the QWERTY keyboard), whereas signal values (bandlimited signals specifically) obey the ordering of real numbers or integers.  The difference is crucial and explains the problems and contradictions with the most common definition of signal or image entropy: replace the state (symbol) probability in Shannon's entropy equation with the probability of a signal (image) value.  Most, if not all signal and image processing textbooks [13,14] use this definition of signal (image) entropy, where the probability is the occurrence of the signal values (image intensities) and the distribution is the unit normalised histogram of those occurences.  This is also the definition supplied with the widely used software MatLab [9].  The problem with this definition is that it is totally blind to any spatial structure. For example a uniformly distributed random signal (image) gives the same entropy value as a gentle sawtooth signal (linear gradient image), yet the former is visibly much more complex than the latter. Although various solutions have been proposed, there seems to be a fundamental problem at the heart of this entropy definition.

Perhaps the second most prevalent interpretation of entropy for signals and images is the Maximum Entropy (MaxEnt) definition, which is successfully deployed in much astronomical imaging. The image itself is assumed to be a probability density function and then used directly in Shannon's entropy equation [15]. This seems intuitively pleasing as it corresponds to the probability of a photon landing on a particular sensor element of an imaging array. It is important to note that this definition is based on signal (image) deconvolution theory [16] and does not suggest that the computed entropy value is related to how many bits might be necessary to non-redundantly encode the signal (image). It transpires that the Shannon entropy form can be replaced by almost any suitable convex function [17,18] and the MaxEnt algorithm works just as well. Bracewell [11] clarifies:

> *The whole philosophical basis of the maximum-entropy method can be put aside in favor of an empirical understanding that is well described by Nityananda and Narayan (1982) and Narayan and Nityananda (1984, 1986). These authors have found satisfactory results in practice with a modified entropy that can have a variety of functional forms including $\sqrt{b}$ and $-\exp(-b)$, provided that, like $\log b$ and $b \log b$, they are convex upward and have a negative third derivative.*

Many other authors have noted the problems and contradictions of these two definitions when applied to images. Various remedies have been suggested. For example Starck [16] reviews the problems with previous definitions of image entropy and then uses a summation over multi-scales to define an image entropy with some desirable properties. In an early and highly influential paper Haralick [19] used a "gradient entropy" to detect regions of visual interest in images and textures. The idea is the basis for Dalal's [20] much cited "histogram of oriented gradients" which has subsequently been widely applied to image feature recognition.

Several authors have returned to thermodynamic/statistical-physics fundamentals to redefine entropy for 2D spatial distributions [21] and patterns [22]. Indeed Feldman and Crutchfield [22–24] have often noted that "there is no natural, unique expression for the excess entropy in two dimensions". However non-redundant encoding and communication has not been an explicit consideration. Kersten [25] considered n-th order conditional entropies for image pixels and their relation to redundancy and the ability to predict the values of missing pixels in natural images. More recently Chandler [26] has developed a method to estimate the true entropy or redundancy of natural images using a novel nearest neighbour technique called the proximity distribution.

Before the advent (in the early 1990s) and overwhelming success of wavelet and DCT (discrete cosine transform) methods for lossy image compression there was a great deal of work on linear predictive image coding. Although much of this work seems to have been forgotten it still underpins most implementations of lossless compression of images. For example the early work of Rice and Plaunt [27] used the (second order Shannon) observation that the best predictor of a pixel value is the previous pixel value in a 1D data stream. Although there are strong objections to using compressibility as a measure of complexity [28], it has also been argued that a working definition of image information content and (the complementary concept of) redundancy is lossless compression file size (in bits, or more usefully, bits-per-pixel). In this context Rabbani [29] has noted:

> *A frequently asked question is how much lossless compression can be achieved for a given image. In light of the noiseless source coding theorem, we know that the bit rate can be made arbitrarily close to the entropy of the source that generated the image. However, a fundamental problem is determining that entropy. An obvious approach to estimating the entropy is to characterize the source using a certain model and then find the entropy with respect to that model.*

The lossless compressed image formats jpg-ls, png and webp all use variations of the classic 2D DPCM (Differential Pulse Code Modulation) linear predictive coding (see Rabbani for extensive discussion). These techniques have retained some early constraints that are no longer generally applicable. For example it is assumed that encoding incurs causality in that the image pixel values above and immediately to the left of a raster scan point are known and that other values are unknown and have yet to be encountered or coded. This enforces a strong horizontal/vertical asymmetry. Nonetheless, except for some special situations or hardware (such as low memory semiconductor chip compressors and very high volume/bandwidth streaming), it is not unusual to have access to all pixel values when image compression commences. We will return to the assumption of causality in the next section. Furthermore, we do not expect a vertical or horizontal image reflection to change the image entropy measure, yet it inevitably changes the lossless compression file size. Likewise, image rotation (assuming appropriate definition of discrete image rotation [30]) invariably changes output file size, which contradicts Shannon (book page 91 [3]) who actually proved that his entropy is invariant to coordinate rotation.

Image compression explicitly based on partial derivatives has been proposed by Ilbery [31,32] although the implementation uses a scanline (viz causal) method not unlike the 2DPCM methods before.

We are aware of at least three other prior approaches that use gradient information to define a measure of image entropy:

- Pluim et al [33] use a gradient measure (actually a magnitude weighted and doubled orientation) to add spatial information to specifically non-spatial mutual information (MI). The idea is to overcome the blindness to spatial variation of Shannon-like mutual information when aligning 3-D tomographic datasets.

- Ramos et al [34] developed an entropic form that can be recognised as the Shannon entropy formula with the probability replaced by the spatial distribution of the gradient itself. This is reminiscent of the aforementioned MaxEnt formulation wherein the spatial distribution of image intensity replaces the probability distribution.

- Shams et al [35] introduce a new mutual information based registration method in which the angular distribution of gradient intensity is used to introduce a spatial dimension to MI.

The above three do not claim to have an entropy measure that corresponds directly with an encoding-decoding scheme that can achieve (or even approaches) non-redundant communication.

This section would not be complete without mentioning Solomonoff-Kolmogorov-Chaitin algorithmic complexity. An assumption of Shannon's information measure is that specific symbol "probabilities" define the entropy and this implies possible algorithms to achieve the optimal symbol coding. But the complementary area of algorithmic complexity theory suggests that the best (least redundant) encoding algorithm is unknown (and in general unknowable) for generic images [36]. The task of computing an algorithmic image information content is then impossible to derive from this theory.

In conclusion to this section it seems that von Neumann's comment *"...no one knows what entropy really is"* accurately describes the current situation with respect to Shannon's information theory applied to signals, and images in particular. We have seen the thermodynamic concept of entropy, defined from the number of micro-states in a macro-state, applied to a wide variety of quite different "micro-state" interpretations.

For the remainder of this paper we eschew the thermodynamics and statistical mechanics interpretation of entropy, instead concentrating on Shannon's original and quite specific intention: defining the amount of non-redundant information needed to communicate a signal or message. As Neal Stephenson observed [37]:

> *All information looks like noise until you break the code.*

# 3  Re-interpreting Shannon's 1D symbol entropy

Shannon's original development (from here we refer to Shannon and Weaver's book [3]) used examples drawn from sequences of English text. The concepts need to be updated a little for the case we are interested in, i.e. bandlimited signals (discrete or continuous). Shannon also discussed the channel capacity of bandlimited systems, but only considered ensembles of signals and in particular the entropy rates for white noise and filtered white noise. In contrast we wish to define the information content of a specific bandlimited signal, rather than some abstract (and unknowable) ensemble of signal classes.

To aid analysis let us consider the concepts in Shannon's theory which are essential and those that seem merely incidental.

ESSENTIAL

- Optimally encode a symbol using code lengths based upon the symbol's (known) likelihood of occurrence.
- Parsimony: known correlations in data can be used to obtain a lower redundancy encoding
- Rotation invariance: a lesser known coordinate transform property (from page 91 section 9 of the book)

INCIDENTAL (non-essential)

- Symbols represent text
  - (we replace this by samples of a bandlimited signal)
- Second order *digram* structure is used denote symbol correlation (or co-occurrence)
  - (we replace this by a derivative or finite difference of samples of a bandlimited signal)
- Symbols or signals arrive in a causal stream
  - (we instead assume our signals have values randomly accessible in memory)

We are now in a position to slightly re-interpret Shannon's entropy in a manner that will allow a natural progression to higher dimensions. In particular, by relaxing the need for a causal structure we avoid introducing asymmetry in 1D, which is one of the principal sources of difficulty in extending to higher dimensions.

## 3.1  Shannon entropy for 1D bandlimited signals

In this initial exposition we consider mostly discrete (summation) formulations, however sometimes we use the continuous or integral formulation if it gives more insight into the underlying processes. The discrete and continuous formulations can both be subsumed into rigorous, single notation using convolution versus Fourier multiplication

Consider a discrete bandlimited signal f containing N samples.

$$f(n), \quad 1 \leq n \leq N \tag{1}$$

We formally define a computable histogram (or probability density function, PDF) generating operation for the raw signal phase space using the Kronecker delta:

$$\rho_i = \frac{1}{N}\sum_{n=1}^{N}\delta_{i,f(n)}, \quad \sum_{i=I_{min}}^{I_{max}}\rho_i = 1, \quad I_{min} = \min\{f\}, I_{max} = \max\{f\} \tag{2}$$

The signal f is typically composed of unsigned 8 or 16 bit integer values. For 8 bits $I_{max}$=255, $I_{min}$=0. The Kronecker delta is a convenient way to describe the discrete binning operation required to generate a 1D histogram $\rho_i$. As we step through index n, the Kronecker delta adds a unit at a histogram value defined by i=f(n), and continues until all N values are binned. The end result is a histogram ρ indexed by i. We hope to avoid Kolmogorov's objection [38] to the use of the term *probabilistic* versus *combinatorial* by explicitly defining the above density (histogram) function from the (known) data itself.

In Shannon's original derivation of the information entropy of a symbol stream the probability of the system being in cell *i* of it phase space (split into *I* cells on total) leads to the conventional histogram entropy rate (in bits per symbol):

$$H(f) = -\sum_{i=1}^{I}\rho_i \log_2 \rho_i \tag{3}$$

We show the discrete form above but note that a continuous integral formulation is possible with suitable constraints on the coordinates. For many signals occurring in real life there is a correlation between nearby signal values and this corresponds to a reduction in the derivative signal's RMS (root mean square) bandwidth. We can rewrite the entropy computation to take advantage of this correlation. If the best predictor of a signal sample value is the preceding sample value [27] then the difference of these sample values is a simple way to characterise decorrelation. Many researchers [29,39] have observed the histogram compression associated with signal derivatives, therefore we consider the PDF p of a derivative or finite difference signal $f_x$(n) with N samples:

$$\begin{aligned}
f_x(n) &= g_x * f = \nabla f(n) \approx f(n) - f(n-1), \\
p_j &= \frac{1}{N-1}\sum_{n=2}^{N}\delta_{j,f_x(n)}, \\
&\sum_{j=J_{min}}^{J_{max}} p_j = 1, \ J_{min} = \min\{f_x\}, \ J_{max} = \max\{f_x\}
\end{aligned} \tag{4}$$

For brevity we call this easily computable density function *deldensity* (see later section 5.2 for justification), and also to separate it from Kolmogorov probabilities and previous incarnations of gradient entropy [19]. Note that deldensity has some features in common with Shannon's symbol transition probability. The Kronecker delta formulation is amenable to Fourier transformation which makes it a tractable quantity in various computations. For an 8 bit signal -255≤ J≤ 255 with a two-sample finite difference.

The corresponding, second order, Shannon information entropy is now simply:

$$H(\nabla f) = H(f_x) = -\sum_{j=J_{min}}^{J_{max}} p_j \log_2 p_j \qquad (5)$$

Again we coin a new term *delentropy,* from the H(∇f) notation above, to allow compact descriptions and to distinguish our new, second order, information measure (see section 5.2).

Shannon outlined a specific scheme (or algorithm) to encode a symbol sequence at the computed entropy rate and subsequently decode. Huffman [40] derived a scheme for achieving the optimal rates predicted by Shannon. Lower redundancy is achieved by using sample correlations. Shannon incorporated sample correlations using *digram, trigram,* and *N-grams* in certain situations and by using transition probabilities in others. We can do the same using our derivative signal. Transition probabilities are explicitly causal (left to right symbol ordering) so instead we use the finite difference or derivative in equation (4). The decoding task then shifts to reconstruction from a signal derivative.

Consider an encoder which generates a mean value and the local derivative estimate. In the following we use a continuous form for ease of exposition but but note that the results hold for the discrete case too. It is possible to frame the derivative operation as a convolution, in which case the convolutional notation covers both continuous and discrete cases:

$$f_x(x) = \frac{\partial f(x)}{\partial x} = \nabla_{1D} f(x) = g_x * f \qquad (6)$$

The continuous domain Fourier transform and inverse are defined :

$$F(u) \equiv \int_{-\infty}^{+\infty} f(x)\exp(-2\pi i u x)\,dx, \quad f(x) \equiv \int_{-\infty}^{+\infty} F(u)\exp(+2\pi i u x)\,du$$
$$F(u) \xleftrightarrow{FT} f(x) \qquad (7)$$

Hence by the derivative and central value theorems we obtain a Fourier multiplier ($G_xF$) form applicable for both continuous and discrete cases:

$$\int_{-\infty}^{+\infty} f(x)\,dx = F(0)$$
$$f_x(x) \xleftrightarrow{FT} F_x(u) = 2\pi i u F(u) = G_x F \qquad (8)$$

Classically a function is recovered from its derivative (within an additive constant) by integration from the left side:

$$f(x) - f(a) = \int_a^x f_x(t)\,dt \qquad (9a)$$

The above is a version of the Fundamental Theorem of Calculus and is the definition of the derivative $f_x$ in terms of an integral. Note how the integration from the left introduces a subtle causal bias in the common form of the fundamental theorem, which is also reflected in the *causal* approach to much signal processing. However the derivative inversion could just as easily have proceeded from the opposite, anti-causal, direction.

$$f(b) - f(x) = \int_x^b f_x(t)dt \tag{9b}$$

$$2f(x) = \int_a^x f_x(t)dt - \int_x^b f_x(t)dt + f(a) + f(b) \tag{10}$$

The above is a symmetric reformulation of the fundamental theorem in which the two endpoints must also be known in order to recover f(x). The formulation is simply the average of the causal and anti-causal forms. In higher dimensions Stokes' Theorem relates a volume integral to a boundary integral. Looking at the 1D problem in the Fourier domain immediately reveals the even-odd symmetry.

$$f(x) \xleftarrow{FT} \frac{F_x(u)}{2\pi i u} = \frac{F_x}{G_x} \xleftarrow{IFT} f_x(x) * \frac{\operatorname{sgn}(x)}{2} = f_x * g_x^{-1} \tag{11}$$

If the derivative corresponds to a Fourier domain multiplication, then the anti-derivative is a Fourier domain division. Fourier division by (πiu) is equivalent to convolution by the signum function. Care has to be taken at the Fourier origin to avoid the zero division, but in the discrete case this is easily achieved by using the F(0) value from equations (8) to replace the Fourier origin (DC) value. Note that a signum is defined to have a singular point at the origin, set equal to zero. This means that when combining a causal and anti-causal signal the central value cancels out. Hence any sample of f can be recovered from the mean and ALL samples of the derivative EXCEPT at the sample itself. This is a form of extreme nonlocality, common to pseudo-differential operators, which Havin calls antilocality [40].

## Loss of Nyquist Frequencies

The continuous Fourier transform treatment above does not include the inevitable periodic structure invoked when sample signals and discrete Fourier transforms are used in practice. For detailed treatment see Brigham [41]. The periodic continuation of an odd function (like u, 1/u, x or sgn(x)) required a sudden change from positive to negative in the vicinity of the higher frequency (u) or largest displacement (x). Then change enforces a zero value at the cross-over point, typically the Nyquist frequency. Note that this effect does not occur for positive, even functions.

## Recapping Our Shannon Reinterpretation

In this section we have seen that a signal can be represented by its derivative which generally has a more compact histogram (PDF) so the corresponding Shannon entropy is reduced. Huffman's coding scheme allows us to encode and decode a stream of symbols representing these derivatives. Encoding proceeds by first forming the finite difference, then Huffman encoding the difference values based on the histogram of finite differences. The mean (or DC) value is computed and stored separately. The first step of decoding simply generates the raw derivatives from the Huffman encoded derivatives. We then use Fourier derivative inversion (or symmetric causal/anti-causal convolution) to recover the bulk of the rawsignal. The mean is then added and the complete original signal is obtained. The aforesaid process resembles 1D DPCM signal encoding/decoding except that we have explicitly utilised a symmetric formulation. The newfound symmetry will allow us to overcome intractable asymmetries that often block attempts to extend analytical methods to higher dimensions

## 3.2 Derivative entropy in 1D and Fourier bandwidth

In equation (4) we defined the histogram deldensity p for the derivative signal $f_x$. The corresponding delentropy follows in equation (5). It turns out that another well known indicator of signal information content – namely the Fourier spectrum - is closely related.

Let us define the RMS width σ (or second moment) of the deldensity. Here we consider discrete signals and compute the first and second moments of deldensity

$$\overline{j} = \sum_{j=J_{min}}^{J_{max}} jp_j, \qquad \overline{i^2} = \sum_{j=J_{min}}^{J_{max}} j^2 p_j \qquad \sigma^2 = \sum_{i=J_{min}}^{J_{max}} (j-\overline{j})^2 p_j \qquad (12)$$

Using the integer sifting property of the Kronecker delta function we can write the corresponding moments of the derivative original signal:

$$\overline{j} = \frac{1}{N-1}\sum_{n=2}^{N} f_x(n) = \overline{f_x}, \quad i^2 = \frac{1}{N-1}\sum_{n=2}^{N} f_x^2(n) = \overline{f_x^2}, \quad \sigma^2 = \frac{1}{N-1}\sum_{n=2}^{N}\left[f_x(n)-\overline{f_x}\right]^2 \qquad (13)$$

Now we introduce the discrete (and recentered) Fourier transform of the signal and its derivative with N=2M+1:

$$F(k) = \frac{1}{2M}\sum_{m=-M}^{M-1} f(m)\exp(-2\pi imk), \quad f(m) = \frac{1}{2M}\sum_{k=-M}^{M-1} F(k)\exp(+2\pi imk)$$
$$(2\pi ik)F(k) = \frac{1}{2M}\sum_{m=-M}^{M-1} f_x(m)\exp(-2\pi imk), \quad f_x(m) = \frac{1}{2M}\sum_{k=-M}^{M-1} (2\pi ik)F(k)\exp(+2\pi imk) \qquad (14)$$

The second line corresponds to the Fourier derivative theorem. The Plancherel-Rayleigh-Parseval [42] theorem relates power in one domain to power in the other (Fourier) domain:

$$\overline{f_x} = 0$$
$$\overline{f_x^2} = (2\pi)^2 \overline{k^2 F^2} \qquad (15)$$
$$\sigma^2 = (2\pi)^2 \overline{k^2 F^2}$$

We can conclude from the last equation that the RMS width σ of the deldensity σ (usually measured in bits) is directly proportional to the mean-square Fourier spectral width of the underlying signal f. In other words the deldensity width mirrors signal bandwidth. The width of the deldensity is also an indicator (albeit via a more complicated logarithmic relation) of the final delentropy value..

## 3.3 Sample correlation, co-occurrence and propinquity

The symmetric formula for sample value reconstruction via convolution given in equation (11) invokes the notion of what we shall call *propinquity* (or Havin's *antilocality* [40]). Any sample value in a signal depends equally on ALL the surrounding sample derivatives, except for the derivative at the chosen sample itself. We can therefore interpret the use of the derivative in signal coding as a deceptively simple but deep way of intertwining the spatial structures at every sample. We suggest that using the derivative is a simple way to invoke the desirable concepts of sample co-occurrence and correlation. As we have seen in section 2 above, (add references to Leibovic, Chandler, Feldman 2D) these concepts often underpin alternative definitions of entropy for signals.

The derivative process is exactly invertible, except for the lost DC (or mean) and Nyquist values (see Appendix A1 for more discussion of lost Nyquist information). The fractional increase in the encoded data size required to store this auxiliary data is small and decreases inversely with the total size of the signal. For example, if we consider a 1D signal familiar to anyone who listens to digital music or podcasts: an audio WAV file. For a typical lossless audio wave signal of 3 minutes saving the DC and Nyquist would require about 36 bits or a 0.00001% increase in the uncompressed file size and entropy. However, in practice, real audio signals have both the DC and Nyquist components removed before any digital processing by analogue filters.

# 4 Extending Shannon's re-interpreted entropy into 2D

It has long been known in optics that Shannon's analysis can be used to define the channel capacity of an imaging system. Felgett and Linfoot calculated the information capacity of an ordinary imaging system [43]. Accounting for noise, time dependence and polarization yields an invariant as shown by Cox and Sheppard [44]. The idea has been extended to 3D imaging using all six electric and magnetic polarization states [45]. However our real interest here is not the channel capacity, but rather the information content of a particular image.

We limit our discussion to grayscale images.

## 4.1 Information in a scalar image

Shannon defines different orders of entropy corresponding to different approximations to English text:

> Zeroth order approximation has all symbols independent and equi-probable,
>
> First order approximation has symbols independent but with different probabilities,
>
> Second order approximation has symbol-pairs (digrams) with known probabilities
>
> Third order approximation has symbol-triplets (trigrams) with known probabilities

Shannon's concepts of zeroth and first order information content already extend into higher dimensions with no problems. The zeroth order as we know communicates messages with no compression. For example each 1-byte pixel is conveyed at a rate of 8 bits per pixel. The first order utilises the intensity histogram in Shannon's classic entropy equation, but often achieves little or no reduction in bit rate for very simple images (like a grey wedge gradually increasing from 0 to 255 over the width if the image, yielding 8 bits per pixel). Both zeroth and first order entropies here are isotropic and rotation invariant, but they are blind to any image structure whatsoever, and rarely exhibit parsimony.

## 4.2 Coding an image from a gradient vector

The question is can we now extend Shannon's second order information measure beyond 1D? We would like to take advantage of the histogram shrinking properties of the derivative (or finite difference) operator we saw in section 3.

As our re-interpreted second-order 1D entropy is based on the 1D signal derivative, extending it to 2D begins with the 2D derivative of the 2D signal. Then the two derivative components of the gradient field of a scalar image can be immediately used in Shannon's definition of joint entropy (page 51 in the book). In a similar manner to equation (5):

$$H(f_x) = -\sum_{i=1}^{I} p_i \log_2 p_i$$
$$H(f_y) = -\sum_{j=1}^{J} p_j \log_2 p_j \qquad (16)$$
$$H(\nabla f) = H(f_x, f_y) = -\sum_{j=1}^{J}\sum_{i=1}^{I} p_{i,j} \log_2 p_{i,j}$$

Here $p_{i,j}$ is a joint probability density function, which can be interpreted as a 2D density function. It can be explicitly computed from the 2D data using the Kronecker delta formalism introduced earlier in equation (4). We assume a 2Mx2N pixel image f(m,n) so that later DFT relations simplify. Note that the 1D densities are obtained by projections of the 2D density.

$$\left. \begin{array}{l} p_{i,j} = \dfrac{1}{4MN} \sum_{n=-N}^{N-1} \sum_{m=-M}^{M-1} \delta_{i,f_x(m,n)} \delta_{j,f_y(m,n)}, \quad \sum_{j=1}^{J} \sum_{i=1}^{I} p_{ij} = 1 \\ p_i = \sum_{j=1}^{J} p_{i,j} = \dfrac{1}{4MN} \sum_{n=-N}^{N-1} \sum_{m=-M}^{M-1} \delta_{i,f_x(m,n)}, \quad \sum_{i=1}^{I} p_i = 1 \\ p_j = \sum_{i=1}^{I} p_{i,j} = \dfrac{1}{4MN} \sum_{n=-N}^{N-1} \sum_{m=-M}^{M-1} \delta_{j,f_y(m,n)}, \quad \sum_{j=1}^{J} p_j = 1 \end{array} \right\} \quad (17)$$

For example the deldensity defined in equation (17) of a continuous image limited within the rectangle -A<x<A, -B<y<B, can be viewed as collection of 2D Dirac deltas:

$$p(\nabla f) = p(f_x, f_y) = p(\xi, \eta) = \frac{1}{4AB} \int_{-B}^{+B} \int_{-A}^{+A} \delta\left(\xi - \frac{\partial f(x,y)}{\partial x}, \eta - \frac{\partial f(x,y)}{\partial y}\right) dxdy \quad (18)$$

Equations (16-18) and their connections with optical focusing, gradient mapping, catastrophe theory, and quantum mechanics are discussed in more detail in Appendix A2.

Shannon's joint entropy (page 51, section 3 of the book) obeys the following inequality with the separate derivative entropies defined in equation (16):

$$H(\nabla f) = H(f_x, f_y) \leq H(f_x) + H(f_y) \quad (19)$$

On the surface this relation appears to demolish any reason to use a gradient-based second-order Shannon entropy. This is because the worst case (namely equality) has the joint entropy $H(f_x,f_y)$ equal to the sum of the separate derivative entropies. This means it would be better (less redundant) to encode the 2D image using just one (x or y) derivative component. Have we missed something?

## 4.3 Papoulis generalized sampling halves the delentropy

At this stage we introduce a concept vital for achieving the lowest redundancy in the vector gradient representation . The idea is based on a refinement of Shannon's bandlimited sampling theorem [7]. Papoulis has shown that if we have multiple versions of a sampled signal, where the sample values are related by a linear filtering, then the signal can be reconstructed at a higher sampling density than that of any of the given versions [6]. The Papoulis Generalized Sampling (PGS) expansion can be extended to higher dimensions. In the case of 2D with two sample values (namely the partial derivatives $f_x$, and $f_y$) we can use two intertwined quincunx lattices each with half the sampling rate of the reconstruction lattice, as shown by Kovacevic [46]. The full details of reconstruction are presented in Appendix A1. The crucial point here is that we can now achieve parsimony because PGS can effectively halve the bit rate of the delentropy $H(f_x,f_y)$:

$$\frac{H(f_x,f_y)}{2} = H_{PGS}(f_x,f_y) = H_{PGS}(\nabla f) \leq \frac{H(f_x)+H(f_y)}{2} \quad (20)$$

The two terms on the Right Hand Side (RHS) correspond to the entropies of the compacted histograms typically encountered in DPCM coding of images [29,39]. Usually the x-derivative is chosen for the convenience of scan-line processing. However an image may have an arbitrary orientation and the structure may be highly directional or tending more to the isotropic. Consider the two extremes and their implications for equation (20). The first case has all the structural variation in the x-direction:

$$H(f_y) = 0, \Rightarrow H_{PGS}(f_x,f_y) = H_{PGS}(\nabla f) \leq \frac{H(f_x)}{2} \quad (20a)$$

The result is that 2D delentropy achieves half the entropy rate of single partial derivative entropy. The second case is for an image with has no overall directional structure (effectively isotropic):

$$H(f_y) = H(f_x), \Rightarrow H_{PGS}(f_x,f_y) = H_{PGS}(\nabla f) \leq H(f_x) = H(f_y) \quad (20b)$$

The delentropy is still less than or equal to either of the single partial-derivative entropies. 2D delentropy achieves the lowest redundancy and parsimony is retained.

Effectively we have taken the localised gradient entropy measure developed by Haralick [19] for texture classification and extended it to measure the information in an entire image. If the original image is considered a 2D scalar potential then the image of interest is the corresponding gradient vector field. The corresponding Jaynes phase space is then the (2D) distribution of vector field values. As we have seen it can also be viewed rather conveniently as a joint entropy with a phase space of two partial derivatives. Two partials are less redundant than one!

## 4.4 Pixel Propinquity, a surprising connectedness

In appendix A1 we show how a bandlimited image can be reconstructed from the gradient component $f_x$ sampled on a half-rate quincunx grid and $f_y$ sampled on the interwined (coset) quincunx grid. The reconstruction uses the inverse gradient operator which has an inverse radial variation when expressed as a convolution kernel. Interested readers can check Arnision's work (ref 2004) and the references therein The relationship is easiest to illustrate in the continuous domain.

$$h(x,y) = Df(x,y) = \left(\frac{\partial}{\partial x} + i\frac{\partial}{\partial y}\right)f(x,y) \xleftrightarrow{FT} 2\pi i(u+iv)F(u,v) = H(u,v)$$

$$h(x,y) = g(x,y) ** f(x,y) \xleftrightarrow{FT} G(u,v)F(u,v) = H(u,v) \quad (21)$$

$$f(x,y) = g^{-1}(x,y) ** h(x,y) \xleftrightarrow{FT} \frac{F(u,v)}{G(u,v)} = F(u,v)$$

$$g^{-1}(x,y) = \frac{e^{-i\theta}}{2\pi r} = \frac{1}{2\pi(x+iy)} \xleftrightarrow{FT} \frac{1}{2\pi i(u+iv)}$$

Note that the final two functions are self-Fourier transforms [47] with many applications in image processing. The central singular value is actually zero by asymmetry. The inverse gradient operator is a convolution with kernel $g^{-1}$. The above reconstruction of an image has a profound interpretation. The value of an image at a particular pixel is dependent on the gradient values at the ALL surrounding pixels (but not the gradient at the pixel itself, i.e. $g^{-1}(0,0)=0$), and the dependency diminishes inversely with increasing distance *r* from that point. In mathematical terms this ist the pseudo-differential interpretation of the inverse gradient operator . There may be a connection here with Shalizi's concept of complexity (avoiding entropy) based on spatio-temporal light cones [48] because the 2D spatial projection of the 3D cone would have precisely this 1/r distribution.

Hence, if we encode and image as a gradient field (vector) image, then the gradient representation automatically incorporates the sought after pixel co-occurrence [21] or proximity distribution [26], (or $m^{th}$ order predictor on pp61 [29]) of recent 2D entropy proposals. Another way to phrase this is to say that the gradient representation embodies *pixel propinquity.*

# 5 Some properties of the 2D deldensity function

In this section we we look at the implications of the entropy equation (16) and the 2D PDF $p_{i,j}$. What is this 2D function that underpins the new Shannon information theory? The 2D distribution of equation (18) points to what Arnol'd (chapter 14 [49]) eloquently describes as *the gradient mapping*:

> *Suppose that we have a smooth function on Euclidean space. Then the gradient mapping is the mapping that associates to each point the gradient of the function at that point. Gradient mappings are a very special class of mappings between spaces of the same dimension.*

These mappings are perhaps better known as Rene Thom's *catastrophes* [50]. Appendix A2 gives more details.

If we take a digital image f(m,n), shown in Fig 1, form two finite difference estimates of the x and y derivatives and then apply the discrete equation (17) we obtain the result shown in Fig 2. Practically implementation requires binning into a 2D histogram.

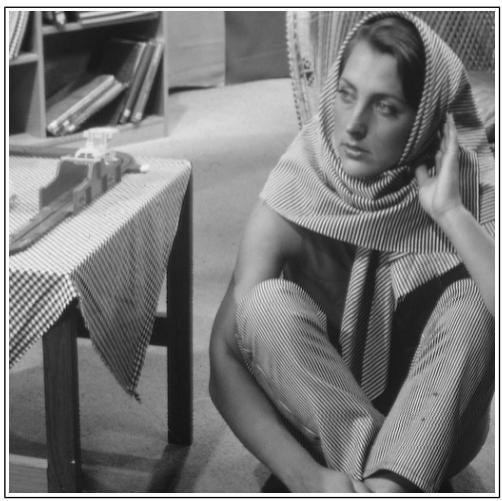

Fig.1,
"Barbara" 512x512 pixel, 8 bit digital image

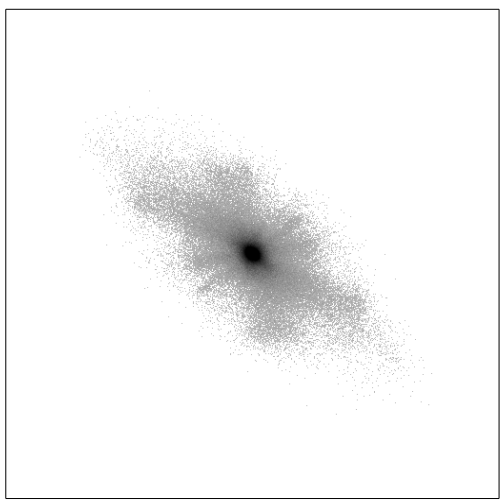

Fig.2
Deldensity $p_{i,j}$ map (511x511) for Barbara. The intensity has been inverted and gamma enhanced to show weak outlying bins. Both the x and y coordinates represent finite differences bin in range ±255 gray-levels

## 5.1 Compute deldensity first

For the direct purpose of entropy estimation and associated gradient coding and decoding of images the deldensity given by equation (17) is generated and then input to equation (16). Note that the logarithm operator requires that all the pixel contributions be added beforehand, so it is not possible to change the summation order in the following; the complete density function must always be computed first:

$$H(f_x, f_y) = -\sum_{j=1}^{J}\sum_{i=1}^{I} \left( \frac{1}{4MN} \sum_{m=-N}^{N-1} \sum_{m=-M}^{M-1} \delta_{i,f_x(m,n)} \delta_{j,f_y(m,n)} \right) \log_2 \left( \frac{1}{4MN} \sum_{m=-N}^{N-1} \sum_{m=-M}^{M-1} \delta_{i,f_x(m,n)} \delta_{j,f_y(m,n)} \right) \quad (22)$$

Actually we should restrict the summation to the image grids specified by PGS. In that case the result in equation (21) is approximately halved.

## 5.2 Del versus Grad nomenclature

Although we use only the gradient mapping in this initial exposition, we conjecture that the del (or nabla) $\nabla$ operator will in general be more widely applicable for other 2D and ND entropy definitions. The del notation can represent gradient $\nabla$, divergence $\nabla \cdot$, or curl $\nabla \times$. For example, the entropy of a 2D vector field can immediately, based on our methodology, be split into the sum of two component entropies; one for the curl-free field, the other for the div free field. In general, given an arbitrary 2D image field, the Helmholtz-Hodge decomposition theorem shows (under specific boundary conditions) that it is the sum of the gradient of a scalar potential and the curl of a vector potential. Implementing Leray projection using forward and inverse Riesz transforms [51] we can isolate two scalar images that generate the full field from the gradient of one and the "curl" of the other. In turn these two components underpin the extremely low entropy phase field coding of fingerprint images [52]. Currently we do not have an unequivocal nomenclature for the new concepts of "2D histogram" and "2D entropy". Possible names for the histogram are: gradient scatterplot, gradient mapping, gradient map, histogram of oriented gradients, gradient caustic, gradient PDF, 2D-PDF, and del-density. None too inspiring. Some possible names for the entropy are: gradient-entropy, gradientropy, gradentropy, and del-entropy. To avoid ambiguity and for brevity we take the liberty of using the neologisms *deldensity* and *delentropy* throughout this paper.

## 5.2 Del computation

There are many ways to estimate the partial derivative components in a discrete image. The most compact possible is perhaps the 2x2 kernel with ±45º finite differences. These can be represented as a complexified kernel (fx +if$_y$) yielding one derivative in the real channel and one in the imaginary, as shown in Fig 3

| +i | +1 |
|---|---|
| -1 | -i |

a

| -1+i | +1+i |
|---|---|
| -1-i | +1-i |

b

| +i | 0 |
|---|---|
| -1-i | +1 |

c

Fig. 3 Three possible discrete gradient kernels a, b and c.

Initially we used the most compact kernel (on the left) with the highest Fourier bandwidth. However, when it comes to Papoulis reconstruction on quincunx sub-lattices the compact kernel exhibits a singularity and doesn't work. So we have rotated the real and imaginary components to obtain the kernel on the right. Although the Fourier bandwidth is slightly lower, this kernel spans both quincunx sub-lattices and reconstruction works as shown in Appendix A1. The kernel is written:

$$f_x(m,n) = [f(m+1,n+1) + f(m+1,n)] - [f(m,n+1) + f(m,n)]$$
$$f_y(m,n) = [f(m+1,n+1) - f(m+1,n)] + [f(m,n+1) - f(m,n)]$$

(23)

For an 8 bit image, the range of values of a two pixel finite difference is in the range ±255, viz 9 bits. Any 8 bit image can then be mapped to a 511x511 deldensity image. Fig.2 and Fig.3 show a typical 8 bit grayscale image and its corresponding deldensity plot. For the purposes of illustration we have used the 2x2 discrete kernel in Fig.1a to compute delentropy in this work. For the kernel in Fig.1b the worst case is would require a 4x255+1=1021 square deldensity image, but high values are actually very unlikely (even for pseudo random noise).

The kernel is convolved with the source image and a complex gradient image produced as in equation (23). For each pixel in the complex target image a binning operation occurs. The smallest deldensity image size may be determined by first finding the maximum of the gradient magnitude before staring the binning operation. The maximum bin count could be as large as the number of pixels in the source image. Typically an INT32 integer value will cover the maximum bin counts encountered. Note we use the double asterisk to denote 2D convolution:

$$f_x(m,n) = g_x(m,n) ** f(m,n)$$
$$f_y(m,n) = g_y(m,n) ** f(m,n)$$

(24)

Implementations can avoids disruptive edge discontinuities by zero padding a one pixel border, or by ignoring the edge computations.

## 5.3 Symmetry and Isotropy

The symmetry and isotropy of the new entropy measure is dependent on the properties of the deldensity which in turn is computed from the discretised del operator. In the continuous case we can make the del operator perfectly isotropic, by which we mean that it is rotationally covariant (i.e. see Farid [53]). Isotropy is easiest to understand in the Fourier domain, and for 2D we take advantage of the complexified derivative operators

$$(g_x + ig_y) ** f = f_x + if_y \xleftrightarrow{FT} (G_x + iG_y)F = Ge^{i\psi}F$$

(25)

For the correct covariance the Fourier magnitude G must be a purely radial function and the Fourier phase ψ must be a purely linear azimuthal angle function. In the case of the discrete 2x2 kernel we find reasonable properties out to about half Nyquist frequency.

## 5.4 Fine details of deldensity plots

The underlying structure of deldensity is not visible in the first example shown in Fig. 2. However if we consider an image with smoother and more gradual gray-level variations, then structure emerges. Perhaps the most striking feature is the appearance of folding in gently curving translucent sheets. Where folds occur the density dramatically increases. For entropy computations the singularities of ρ itself are ameliorated by the use of integrals and summations and normalization. The continuous formulation in equation (18) implies that high densities are generated from image regions with slowly varying gradient. These correspond to regions of the source image that have a vanishing Hessian. The maths is very much like the asymptotic stationary phase approximation found in coherent optical imaging and interferometry [54]. This is not surprising as equation (18) essentially defines the continuous limit of finite ray-tracing through a lens with a thickness profile proportional to the image value f(x,y). Appendix A2 contains a more detailed discussion of these issues.

Consider Fig. 4 which shows an image derived from the image in Fig. 2 by simple low pass filtering and up-sampling to 2024x2048 pixels. The corresponding deldensity shown in Fig. 5 in which a profusion of caustic structures are visible. The important point here is that the deldensity captures the underlying image structure and complexity. Areas of high density result in a low entropy contribution. These in turn correspond to regions of very slowly changing gradient in the source image; regions that can be encoded by a short symbol.

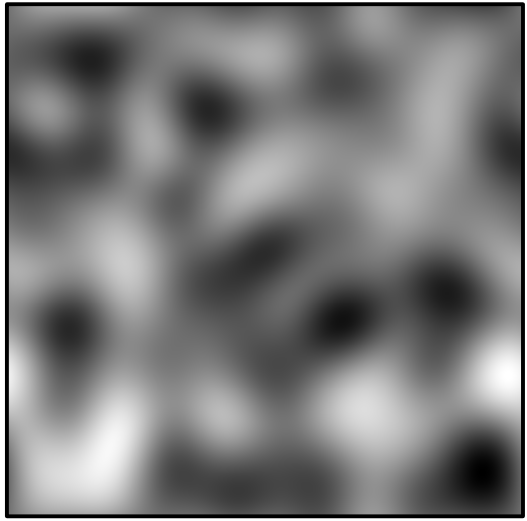

Fig.4
A 2048x2048 pixel 16 bit small bandwidth image:
a low passed version of Fig.1

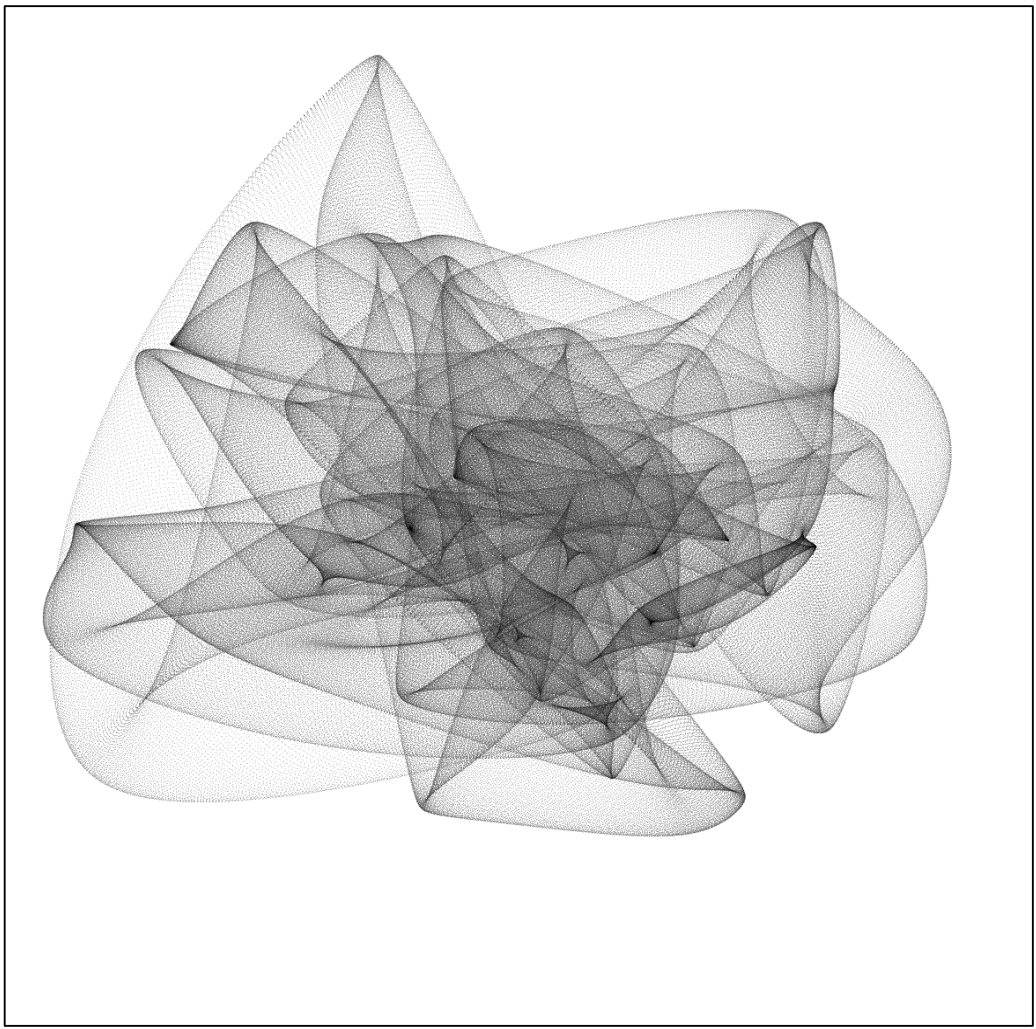

Fig.5
Corresponding 1167x1167 deldensity map for image in Fig.4. The intensity has been inverted and nonlinearly enhanced to show fine structure. The centred x-y coordinates represent differences in the range ±580 gray-levels

## 5.5 Deldensity and 2D Fourier bandwidth

The results of section 3.2 readily extend into 2D. Consider the three second moments

$$\bar{i} = \sum_{j=1}^{J}\sum_{i=1}^{I} i p_{i,j}, \qquad \overline{i^2} = \sum_{j=1}^{J}\sum_{i=1}^{I} i^2 p_{i,j} \qquad \sigma_{\xi\xi}^2 = \sum_{j=1}^{J}\sum_{i=1}^{I} (i-\bar{i})^2 p_{i,j}$$

$$\bar{j} = \sum_{j=1}^{J}\sum_{i=1}^{I} j p_{i,j}, \qquad \overline{j^2} = \sum_{j=1}^{J}\sum_{i=1}^{I} j^2 p_{i,j} \qquad \sigma_{\eta\eta}^2 = \sum_{j=1}^{J}\sum_{i=1}^{I} (j-\bar{j})^2 p_{i,j} \tag{26}$$

$$\sigma_{\xi\eta}^2 = \sum_{j=1}^{J}\sum_{i=1}^{I} (i-\bar{i})(j-\bar{j}) p_{i,j}$$

Again using the integer sifting property of the Kronecker delta function:

$$\bar{i} = \frac{1}{4MN}\sum_{m=-N}^{N-1}\sum_{m=-M}^{M-1} f_x(m,n) = \overline{f_x}, \; \overline{i^2} = \frac{1}{4MN}\sum_{m=-N}^{N-1}\sum_{m=-M}^{M-1} f_x^2(m,n) = \overline{f_x^2},$$

$$\bar{j} = \frac{1}{4MN}\sum_{m=-N}^{N-1}\sum_{m=-M}^{M-1} f_y(m,n) = \overline{f_x}, \; \overline{j^2} = \frac{1}{4MN}\sum_{m=-N}^{N-1}\sum_{m=-M}^{M-1} f_y^2(m,n) = \overline{f_y^2}, \tag{27}$$

$$\overline{ij} = \frac{1}{4MN}\sum_{m=-N}^{N-1}\sum_{m=-M}^{M-1} f_x(m,n) f_x(m,n) = \overline{f_x f_y}$$

$$\sigma_{\xi\xi}^2 = \frac{1}{4MN}\sum_{m=-N}^{N-1}\sum_{m=-M}^{M-1} \left[f_x(m,n) - \overline{f_x}\right]^2, \; \sigma_{\eta\eta}^2 = \frac{1}{4MN}\sum_{m=-N}^{N-1}\sum_{m=-M}^{M-1} \left[f_y(m,n) - \overline{f_y}\right]^2$$

$$\sigma_{\xi\eta}^2 = \frac{1}{4MN}\sum_{m=-N}^{N-1}\sum_{m=-M}^{M-1} \left[f_x(m,n) - \overline{f_x}\right]\left[f_y(m,n) - \overline{f_y}\right]$$

Now we introduce the discrete (and centered) Fourier transform of the signal

$$F(k,l) = \frac{1}{4MN}\sum_{n=-N}^{N-1}\sum_{m=-M}^{M-1} f(m,n)\exp\left[-2\pi i(mk+nl)\right]$$

$$f(m,n) = \frac{1}{4MN}\sum_{l=-N}^{N-1}\sum_{k=-M}^{M-1} F(k,l)\exp\left[+2\pi i(mk+nl)\right] \tag{28}$$

$$f_x(m,n) = \frac{1}{4MN}\sum_{l=-N}^{N-1}\sum_{k=-M}^{M-1} (2\pi ik) F(k,l)\exp\left[+2\pi i(mk+nl)\right]$$

$$f_y(m,n) = \frac{1}{4MN}\sum_{l=-N}^{N-1}\sum_{k=-M}^{M-1} (2\pi il) F(k,l)\exp\left[+2\pi i(mk+nl)\right]$$

The Plancherel-Parseval power theorem means that

$$\overline{f_x} = 0, \; \overline{f_y} = 0 \tag{29}$$

$$\overline{f_x^2} = (2\pi)^2 \overline{k^2 F^2}, \; \overline{f_y^2} = (2\pi)^2 \overline{l^2 F^2}, \; \overline{f_x f_y} = (2\pi)^2 \overline{lkF^2}$$

$$\sigma_{\xi\xi}^2 = (2\pi)^2 \overline{k^2 F^2}, \; \sigma_{\xi\xi}^2 = (2\pi)^2 \overline{l^2 F^2}, \; \sigma_{\xi\eta}^2 = (2\pi)^2 \overline{lkF^2}$$

In other words, all three second moments of the deldensity correspond directly with the Fourier spectrum second moments. It almost seems as if the deldensity foreshadows the Fourier spectrum. It turns out that the connection can be formalised using the asymptotic method of stationary phase. The lens analogy in section 5.3 suggests that the point spread function obtained by ray-tracing can approximated by wave propagation, in which case the amplitude point spread function is just a scaled Fourier transform of a pure phasor function with phase proportional to f(x,y).

$$a(\xi,\eta) \propto \int_{-B}^{+B}\int_{-A}^{+A} \exp\left[i\gamma f(x,y)\right]\exp\left[2\pi i\alpha(\xi x,\eta y)\right]dxdy \tag{30}$$

The deldensity is then the zero wavelength (infinite frequency γ→∞) limit of the intensity point spread function |a|. The Hessian of f determines the origin of caustics and corresponding singularities of the stationary phase approximation of equation (30).

For a given RMS width σ the density shape that gives a maximum value of entropy is a Gaussian. This applies in 2D as well. So we can write an inequality for the maximum value of entropy associated with the three second moments. Kolmogorov considered a similar case for the joint entropy of continuous signals [55].

## 5.6 Deldensity support and entropy limits

If the maximum value of |∇f|=τ, then the delentropy must satisfy

$$H(f_x, f_y) \leq \log_2(\pi\tau^2) = \log_2\pi + 2\log_2\tau$$
(31)

And the maximum occurs where the deldensity is uniformly distributed over a circle of radius τ. For a typical 8 bit image the maximum finite difference requires 9 bits (τ=510), hence H≤19.65 bpp or 9.83 bpp taking PGS into account.

## 5.7 Extension to higher orders

The symmetric form of del can be extended to orders beyond the second simply by recursion. A similar idea was used by Köthe and Felsberg [56], Larkin [57], and Salzenstein [58] for extending the energy operator isotropically into 2D and more dimensions. In 2D Felsberg derived the Hessian 2x2 tensor form from repeated application of the del operator:

$$\mathbf{H}f = \nabla\nabla^T f = \begin{pmatrix} \dfrac{\partial^2 f}{\partial x^2} & \dfrac{\partial^2 f}{\partial x\partial y} \\ \dfrac{\partial^2 f}{\partial y\partial x} & \dfrac{\partial^2 f}{\partial y^2} \end{pmatrix} \tag{32}$$

The Hessian has four components (two of which are equal, for continuous functions) and we conjecture that PGS will allow reconstruction from samples on four intertwined square (rather than quincunx) grids.

How do we extend beyond the third order (Hessian) entropy? Symmetry will be an important consideration. But there are also other important considerations. For example the possibility of defining the entropy for vector valued functions such as vector fields, but also implicitly vector quantities like phase see appendix A3 for more details). We have already demonstrated that decomposing an intrinsically 1D (i1D [59]) image into modulated amplitude and phase allows a huge reduction in redundant information in fingerprint images [52]. The important point is that a vector field can be uniquely decomposed as the sum of a curl-free and a divergence-free field. Each of these fields has a second order entropy that is very low. This suggests that a vector field Shannon-like entropy can be defined and uniquely separated into two entropy terms. We suggest this is an interesting direction for future research.

# 6 Practical Implementation

Once the deldensity binning is complete (equation (17), the delentropy computation can proceed (equation (16)). In practice we only need to bin the x-derivatives on a quincunx sub-lattice and the y-derivatives on the alternate quincunx lattice (see appendix A1). The computed delentropy can use bin coordinates defined by the x-quincunx value and the immediately adjacent y-quincunx value to the right. This choice is arbitrary but needs to be consistently applied to allow perfect decoding latter.

In principle it seems that delentropy be still be 1bpp greater than the old histogram entropy. However it seems unlikely that the worst case deldensity (uniform probability $p_{i,j}$ of all gradients occurring) can happen at the same time as the worst case histogram (uniform distribution of intensity over full range K). For typical images $H_{PGS}(f_x,f_y)$ is much lower than $H(f)$ because adjacent pixels are highly correlated and thus differential images have highly concentrated histograms and deldensities.

Computation of delentropy is actually quite direct and straightforward. No iterations or recursions are required. The computational complexity is linear in the number of pixels of the source image. Several approximations are possible. The simplest applies the chosen del operator (typically grad) to each pixel and computes a deldensity containing every binned pixel. The raw delentropy is computed and the result halved to allow for Papoulis PGS. Another approach gives the exact encoding bit rate by mirroring the actual quincunx sampling of two del components separately. The difference between methods is expected to be small, but may be significant in subsequent research.

Encoding requires explicit computation of the deldensity to guide the Huffman encoding of two partial differential images (see appendix A1 for details).

Decoding requires simple Huffman decoding and discrete Fourier transform inversion giving an overall NlogN computational complexity in the number of pixels. For more details see appendix A1.

We have provided a link to the software used to generate deldensity and delentropy for digitals images [60]. We have also included a link to the Fourier image processing software used to implement our reconstruction from quincunx sampled derivatives [61].

# 7  Comparison of Delentropy and Lossless Compression

Can the new delentropy measure achieve the parsimony we desire?  We have compared delentropy with a number of approaches outlined in Section 2.  In particular we have computed:

- The intensity histogram entropy
- Various lossless image coding entropies computed from the compressed file size:
    - BMP (no compression)
    - GIF (Lempel–Ziv–Welch  encoding)
    - JPG-LS (2D DPCM)
    - JP2000 lossless (reversible integer wavelet transform)
    - PNG (2D DPCM and  Lempel-Ziv-77 + Huffman encoding )
    - WEBP (modern, highly optimised 2D DPCM with LZ77-Huffman encoding)
- The 2D delentropy

The computations have been applied to a small selection of well known image types.  In future it may be informative to apply to a much larger database of images.

Several lossless image coding standards (e.g. PNG and WEBP) utilize predictive coding founded on causal gradient estimation (viz using prior pixel values, above and to the left in scan-line processing).  Other image coding formats (e.g. GIF and PCX) use run length encoding of scan-lines left to right (i.e. causal).  The causal structure of compression method means that 90º or 180º rotation of an image before compression can produce a quite different file size.

There has been quite in change in the perception of signal processing since 1948 when Shannon first proposed his information measure, especially with regard to computer memory.  Although some hardware applications are restricted to scan-line processing (or streaming) and the ubiquitous causal constraints apply, most modern signal and image compression software applications hold the entire signal or image in memory before the processing commences.  Shannon was actually aware of this when he wrote (page 62, book):

> *In general, ideal or nearly ideal encoding requires a long delay in the transmitter and receiver.  In the noiseless case which we have been considering, the main function of this delay is to allow reasonably good matching of probabilities to corresponding lengths of sequences.*

Although it now seems reasonable to ignore causal constraints for many 1D and 2D signal types, the causal constraint re-emerges for very large files like those based on the recently developed 4K video standard.

We have computed the effective overall bit rates for each of the lossless compressors.  The particular values shown correspond to the files compressed by PaintShop Pro X5 (Corel Corp. 2012).  Slightly different values can be expected for different software implementations.

Fig. 6 shows the results

| Description | Image | Lossless Compression [bits per pixel] | | 1-D intensity histogram entropy | 2-D gradient histogram entropy |
|---|---|---|---|---|---|
| **Uniform pseudo-random noise** 8 bit pixels intensity (0-255) 256x256 pixels 65536 bytes | 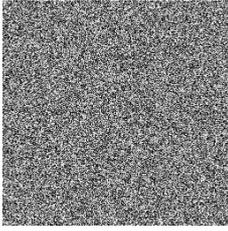 | JPG-LS | 14.2 | 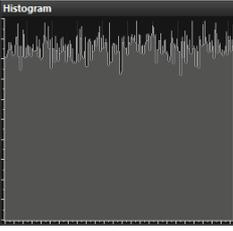 7.99 bpp | 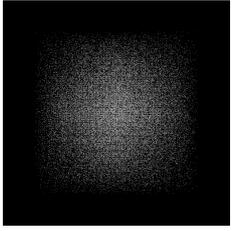 7.59 bpp |
| | | JP2000 | 8.6 | | |
| | | GIF | 6.7 | | |
| | | PNG | 7.9 | | |
| | | WEBP | 5.6 | | |
| **Uniform histogram structured noise** 8 bit pixels intensity (0-255) 256x256 pixels 65536 bytes | 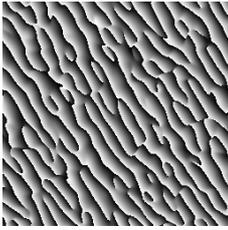 | JPG-LS | 10.7 | 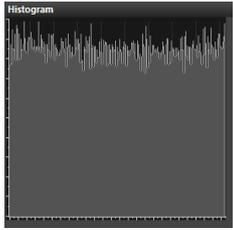 7.99 bpp | 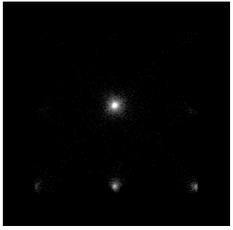 4.40 bpp |
| | | JP2000 | 4.7 | | |
| | | GIF | 8.1 | | |
| | | PNG | 3.9 | | |
| | | WEBP | 2.5 | | |
| **Smooth gradient grey** 8 bit pixels intensity (0-255) 256x256 pixels 65536 bytes | 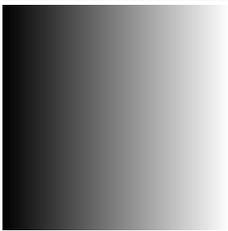 | JPG-LS | 1.2 | 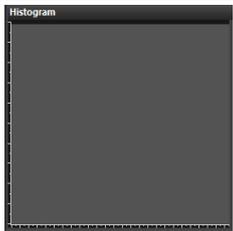 8.00 bpp exactly | 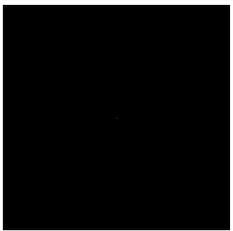 0.00 bpp exactly |
| | | JP2000 | 0.2 | | |
| | | GIF | 1.7 | | |
| | | PNG | 0.1 | | |
| | | WEBP | 0.1 | | |
| **Photo - Barbara** 8 bit pixels intensity (0-255) 256x256 pixels 65 KB | 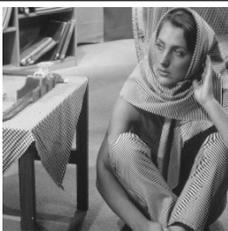 | JPG-LS | 7.6 | 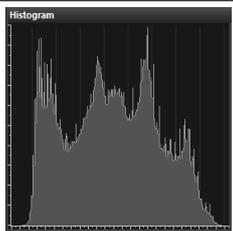 7.63 bpp | 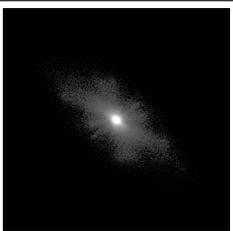 6.05 bpp |
| | | JP2000 | 5.5 | | |
| | | GIF | 9.2 | | |
| | | PNG | 6.2 | | |
| | | WEBP | 3.0 | | |
| **Low pass Barbara image** 8 bit pixels intensity (0-255) 256x256 pixels 65 KB | 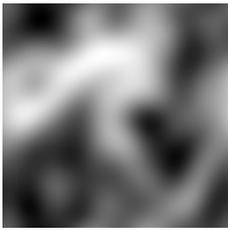 | JPG-LS | 2.4 | 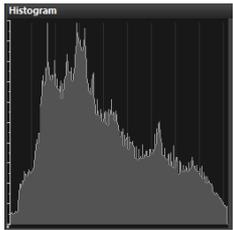 7.79 bpp | 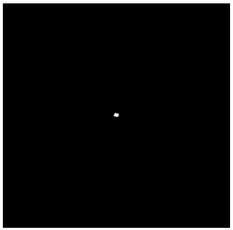 3.41 bpp |
| | | JP2000 | 1.3 | | |
| | | GIF | 5.3 | | |
| | | PNG | 1.8 | | |
| | | WEBP | 0.4 | | |
| **Low pass smooth image Upsampled** 16 bit pixels intensity (0-65536) 1024x1024 pixels 2.05 MB | 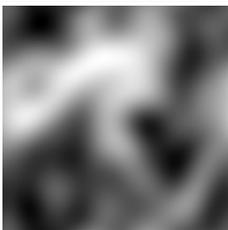 | JPG-LS | NA | 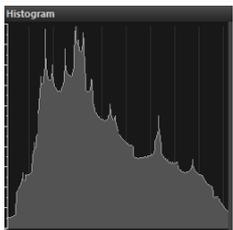 15.76 bpp | 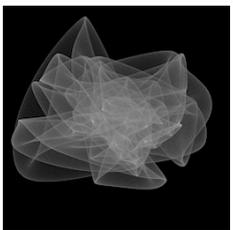 9.11 bpp |
| | | JP2000 | 4.7 | | |
| | | GIF | 8.1 | | |
| | | PNG | 3.9 | | |
| | | WEBP | 2.5 | | |

Fig. 6, Comparison of entropy measures for selection of images

The first notable point about the six chosen images is that the conventional (intensity histogram) entropy is near a maximum in all cases. For the first five it is near 8 bits per pixel, and for the last it is almost 16 bpp.

The delentropy values are always much lower than the conventional entropy, except perhaps for pseudo-random noise in the first image. Delentropy usually falls midway between the GIF and PNG compression rates. In one extreme case (the linear gray wedge) of the third image delentropy has a value of zero because there is only one gradient value for all pixels; deldensity is a single peak.

Generally the WEBP image compression reaches lower bit-rates. However WEBP combines DPCM with a higher order encoding using LZ77. Our objective is to show that second order delentropy yields a value in keeping with the visual perception of image complexity, and not necessarily the very lowest bit rate.

Delentropy uses nothing more than the phase space collapse (deldensity) induced by the del operator to obtain low entropy values.

More testing on a wide range of image types could be informative, but our primary aim here is to initially show how Shannon entropy can be extended to reflect some structural complexity.

# 8  Discussion and future research

We have proposed a way to reconcile Shannon's original concept of information-entropy with the known properties of two dimensional bandlimited images. In particular we have shown how second order properties like pixel co-occurrence and proximity are easily invoked using the vector field formalism for scalar images. Even in 1D we can now interpret Shannon second order entropy with time reflection symmetry.

The desired properties of symmetry, and isotropy emerge naturally from del operators such a grad and curl applied to an underlying scalar potential image. We believe that the del operators can be applied recursively to obtain higher order entropies with similar symmetry and invariance properties. Higher orders such as div-grad (Laplacian), curl-curl and Hessian may have diverse applications.

Parsimony has been achieved by combining two of Shannon's great insights: information theory and bandlimited sampling theory. Without the two-fold efficiency of Papoulis' generalized sampling expansion the vector field entropy would be greater than the scalar entropy in the worst case. In the best cases vector based delentropy is far lower than the old scalar entropy.

The concepts of deldensity and delentropy open up new possibilities of highly tractable mathematics. For example the tomographic properties of 2D deledensity evoke Fourier projection-slice relations and classic Shannon joint entropy inequalities. The del formalism may have implication for research in compressive sensing and sparsity because the grad operator is implicated in total variation reconstructions in 2D and beyond. If we can speculate even further then the Dirac operator or vector derivative formalism used in Clifford algebra has great potential to combine all the different del operators into a single, unified, operator with all the necessary symmetries implicit.

We have not considered video sequences, however the del operator can be extended into the third dimension even if it is temporal rather than spatial. The concepts and mathematical tools easily extend into 3D. In fact the spatio-temporal derivative can mimic aspects of the optical flow method for video compression of coherent motion in video sequences.

It should be mentioned that the mathematical theory of potentials and vector fields already contains many of the techniques we have outlined for reconstruction. The classical Newtonian potential has exactly the same inverse radial form as our reconstruction kernels. The theory has already been generalized to N dimensions.

## Acknowledgements

This paper could not have been completed without Raul Vera's steady encouragement, questioning and support writing the code to compute deldensity and delentropy. I am indebted to my former colleague Peter A. Fletcher at *Canon Information Systems Research Australia* for introducing me to the elegant structure and curious mathematics of high resolution gradient mappings in 2002. I would like to thank Jeffrey A. Hogan at *The University of Newcastle* for suggesting the quincunx down-sampling needed for the reconstruction from undersampled partial derivative images. Michael A. Oldfield wrote the Eigenbrötler software for my prototyping of fast Fourier techniques (such as reconstruction) essential for this research. Peter Tuthill at the Sydney Institute For Astronomy (SIFA) answered my early questions about the meaning of entropy in the Maximum Entropy reconstruction of astronomical images.

# Appendix A1: Image reconstruction from half-rate partial derivatives

When we extended Shannon's information measure to a gradient vector field representation of a scalar potential image we doubled the worst case bit rate.  the new theory would then have a second order entropy potentially greater than the first order were it not for the fact that we are working with discrete bandlimited images.  Each vector component value in a vector field can be considered a linear filtered estimate of the underlying scalar value.  This is precisely the case Papoulis [6] consider when he generalized Shannon's original sampling theory [7] to situations where multiple linear systems are sampled.  In particular a scalar image can be recovered from two vector components of the gradient field, each sampled at half the scalar image sampling rate.  In 2D half-rate sampling is enabled by two intertwined quincunx lattices.  However this particular case has not been considered in the literature previously, so we show how the reconstruction proceeds.

   Although Papoulis originally developed the sampling expansion in 1D, Kovacevic [46] conveniently extended the theory for two half-rate channels in 2D.  Consider the related two channel system of Feilner [62].  We proceed by separating our gradient (or more generally a del) operator into two partial derivatives.  The following analysis is valid for discrete of continuous Fourier transforms.  Lower case functions are general spatial domain and capitalization indicates the corresponding Fourier transform

First consider a discrete approximation of the derivatives (using two or more pixel values) as a convolution with 2D spatial kernels $g_x$ an $g_y$:

$$f_x(x,y) = g_x(x,y) ** f(x,y), \quad f_y(x,y) = g_y(x,y) ** f(x,y) \tag{A1-1}$$

The Fourier multiplier version follows immediately:

$$F_x(u,v) = G_x(u,v) F(u,v), \quad F_y(u,v) = G_y(u,v) F(u,v) \tag{A1-2}$$

 The objective then is to show that the perfect reconstruction filter-bank works with the x-derivative transform in one channel and the y-derivative transform in the other channel.  Apart from two DC and two Nyquist components.

Consider two intertwining quincunx lattices $q_1$  an d $q_2$ composing the usual square lattice $q_0$.  The square lattice has unit spacing.

$$\left. \begin{aligned} q_0(x,y) &= \sum_{j=-\infty}^{+\infty} \sum_{i=-\infty}^{+\infty} \delta(x-i).\delta(y-j) \\ q_1(x,y) &= \frac{1}{2} \sum_{j=-\infty}^{+\infty} \sum_{i=-\infty}^{+\infty} \{1+\cos(\pi[x+y])\} \delta(x-i).\delta(y-j) \\ q_2(x,y) &= \frac{1}{2} \sum_{j=-\infty}^{+\infty} \sum_{i=-\infty}^{+\infty} \{1-\cos(\pi[x+y])\} \delta(x-i).\delta(y-j) \end{aligned} \right\} \tag{A1-3}$$

Taking the Fourier transforms we find quincunx sampling introduces an overlapping, replicated sidelobe with a positive or negative sign:

$$Q_0(u,v) = \delta(u,v)$$
$$Q_1(u,v) = \delta(u,v) + \delta(u-u_0, v-v_0)$$
$$Q_2(u,v) = \delta(u,v) - \delta(u-u_0, v-v_0)$$

(A1-4)

Next we compare the original image sampled at full rate $s_0$, with two partial derivative images sampled at half-rate on the intertwining quincunx grids $s_1$, $s_2$:

$$s_0 = f(x,y)q_0(x,y) \xleftrightarrow{FT} G_x(u,v)F(u,v) = S_0$$
$$s_1 = f_x(x,y)q_1(x,y) \xleftrightarrow{FT} G_x(u,v)F(u,v) + G_x(u-u_0,v-v_0)F(u-u_0,v-v_0) = S_1$$
$$s_2 = f_y(x,y)q_2(x,y) \xleftrightarrow{FT} G_y(u,v)F(u,v) - G_y(u-u_0,v-v_0)F(u-u_0,v-v_0) = S_2$$

(A1-5)

To reconstruct the underlying scalar image f we solve a simple simultaneous equation in the Fourier domain, which amounts to cancelling the duplicated spectrum and then renormalizing the remaining spectrum:

$$\left. \begin{array}{l} S_1 = G_x F + \tilde{G}_x \tilde{F} \\ S_2 = G_y F - \tilde{G}_y \tilde{F} \end{array} \right\}$$

$$\left. \begin{array}{l} S_1 \tilde{G}_y = \tilde{G}_y G_x F + \tilde{G}_y \tilde{G}_x \tilde{F} \\ S_2 \tilde{G}_x = \tilde{G}_x G_y F - \tilde{G}_x \tilde{G}_y \tilde{F} \end{array} \right\}$$

(A1-6)

$$F = \frac{S_1 \tilde{G}_y + S_2 \tilde{G}_x}{\tilde{G}_y G_x + \tilde{G}_x G_y} = \frac{(S_1 \tilde{G}_y + S_2 \tilde{G}_x)}{\Gamma} = (S_1 \tilde{G}_y + S_2 \tilde{G}_x) H, \; \Gamma = \tilde{G}_y G_x + \tilde{G}_x G_y = \frac{1}{H}$$

$$f = (s_1 ** \tilde{g}_y + s_2 ** \tilde{g}_x) ** h$$

(A1-7)

The solution can be written in the Fourier or spatial domain. However it is much easier to compute and control division singularities in the discrete Fourier domain. The Fourier multipliers $G_x$, and $G_y$ typically have zeros at DC and Nyquist frequencies.

Because of the sampling and replication properties of discrete sampled bandlimited quantities on a regular grid (see chapter 6 of Brigham [41]) the discrete Fourier transform (DFT) of the filters simply repeats outside the Nyquist zone. This means that a odd function (like the sine resulting from finite differences) must have a discontinuity as it abruptly changes from positive to negative as it crosses the Nyquist zone. Typically $G_x$ has a line of zeros at u=0 (DC) and v=0.5 (Nyquist), and $G_y$ has a line of zeros at v=0 and u=0.5.

We consider full arrays with an even number of samples in the x and the y directions. This means that the quincunx arrays have the same number of samples. Arrays with an even number of samples this discontinuity produces a midway value of zero at the Nyquist frequency. The easiest way to

compute the shifted multipliers is to apply the DFT (usually an FFT) with modulation in the spatial domain:

$$g_x(u,v)\exp[i\pi(u+v)] \xleftrightarrow{FT} \tilde{G}_x(u,v)$$
$$g_y(u,v)\exp[i\pi(u+v)] \xleftrightarrow{FT} \tilde{G}_y(u,v)$$
(A1-8)

This is the same as the ±1 chequerboard modulation used to recentre raw DFTs. Now we know all the multipliers we can compute the overall filters. Returning to equation (A1-7) we remove the singularities before the division operation:

$$F = S_1 G_1 + S_2 G_2$$
$$\Gamma = 0 \Rightarrow G_1 = G_2 = 0$$
$$\Gamma \neq 0 \Rightarrow \left\{ \begin{array}{l} G_1 = \dfrac{\tilde{G}_y}{\Gamma} \\ G_2 = \dfrac{\tilde{G}_x}{\Gamma} \end{array} \right.$$
(A1-9)

The lost DC and Nyquist lines in x and y need to be recovered. These cannot be recovered by the complex division device [51,63] used in fully sampled differential images because quincunx sampling intrinsically loses information in these regions.

To recover the DC and Nyquist frequencies we add a step in the processing. The simplest method is to store the mean value (DC) of each row and column of the image. The mean needs to calculated to better than half a bit (e.g. a quarter of a bit) to allow the ultimate recovery of each row and column to the nearest bit. Nyquist components are recovered in a similar way for each row and column, except the summation is replaced by a summation with alternating ±1 modulation. After normalisation each Nyquist value needs to be accurate to a quarter bit for perfect recovery.

Fig A1-1 shows how an image is decomposed as two intertwined quincunx sample lattices of partial derivatives. A small addition amount of information is needed to represent the DC and Nyquist components lost by the quincunx sampling.

## Encoding

- The image gradient is computed and only the values of the x and y partial derivatives on alternating quincunx grids (set and coset) $q_1$ and $q_2$ are used.
- 2D deldensity for data is computed.
- Huffman coding of the data is based on deldensity, compressed data is saved
- Two DC and two Nyquist lines are computed and Huffman coded based on line statistics

## Decoding

- Main compressed data is Huffman decoded

- Two separate quincunx grids are constructed on the full grid with zeros inserted as defined by equation (A1-3)
- Array $s_1$ contains the x-derivative as defined by (A1-1)
- Array $s_2$ contains the y-derivative as defined by (A1-1)
- DFT domain filters are constructed from $g_x$ and $g_y$
- Shifted DFT domain filters are constructed from modulated $g_x$ and $g_y$
- Fourier transform is reconstructed based on equation (A1-9)
- Inverse transform gives $f_0$
- Subsidiary compressed (DC-Nyquist) data is Huffman decoded
- DC and Nyquist added back into image resulting in f.

## Algorithm flowcharts

Fig. A1-1 shows the encoding operation, Fig. A1-2 the encoding operation.

## Additional Notes

Like the 1D case in section 3, adding the DC-Nyquist encoding only changes the computed delentropy by a few percent at most. For example consider a one mega-pixel, 8 bit image. The DC and Nyquist rows and columns will contain 4k pixels that need to be stored at 8+2 bit depth. This is a total of 40k bits before compression, or 0.5% of the original file size. Only in extremely low delentropy images will the DC-Nyquist side-channel have a noticeable effect.

Every reconstructed pixel value depends (almost exactly) isotropically on all the surrounding gradient values (pixel propinquity). In the full sampling case the reconstruction is equivalent to a spatial convolution with the inverse gradient operator kernel; a kernel with a reciprocal distance drop-off.

The equations defining G and H for perfect reconstruction have a quite different form than those of Feilner [62] because the del operator invokes not a highpass-lowpass reflection symmetry but rather two antisymmetric ramps. For example the highpass-lowpass approach circumvent the DC-Nyquist problem we encounter.

Another important point is that it seems the most compact derivative kernel, the 2x2 45º is not invertible by the usual two channel system. There is a singularity in the inversion matrix. However, the full Fourier interpolated derivatives obtained by Fig. A1-1 is invertible, but the pixels values on the downsampled quincunx grids (blue dots on diagram) are floating point values that need to be quantized before effective Huffman coding occurs. For lossless coding there may be a 1 bpp penalty incurred. We have not pursued the issue further in this initial exposition.

We note that recently Patel [64] and Sakhaee [65] explored similar gradient (curl-free) reconstruction for Compressed Sensing applications, especially where the image gradient is sparse.

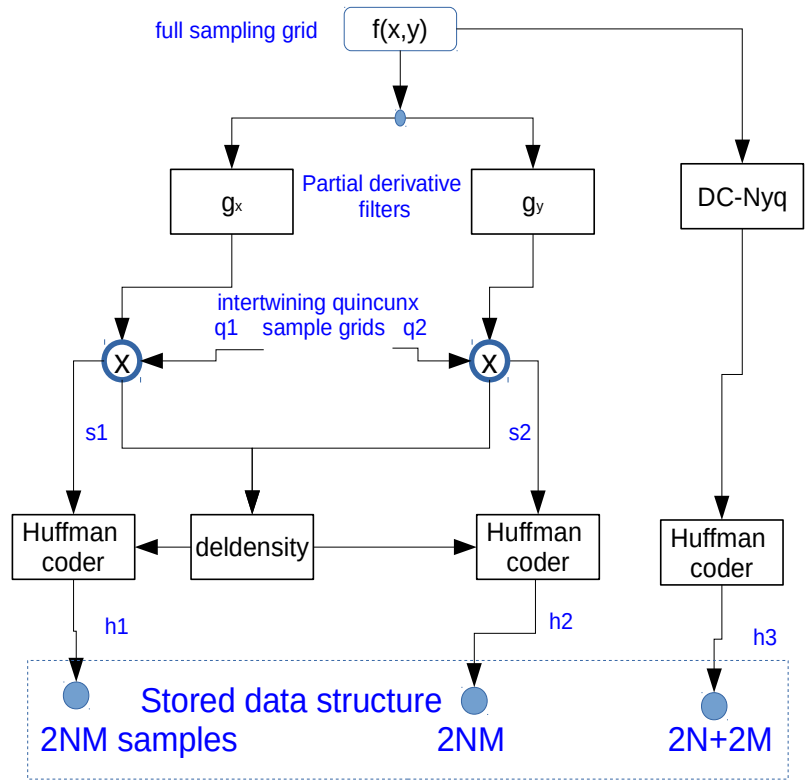

Fig. A1-1
Image encoding as two intertwined partial derivative sample grids and a DC-Nyquist side-channel

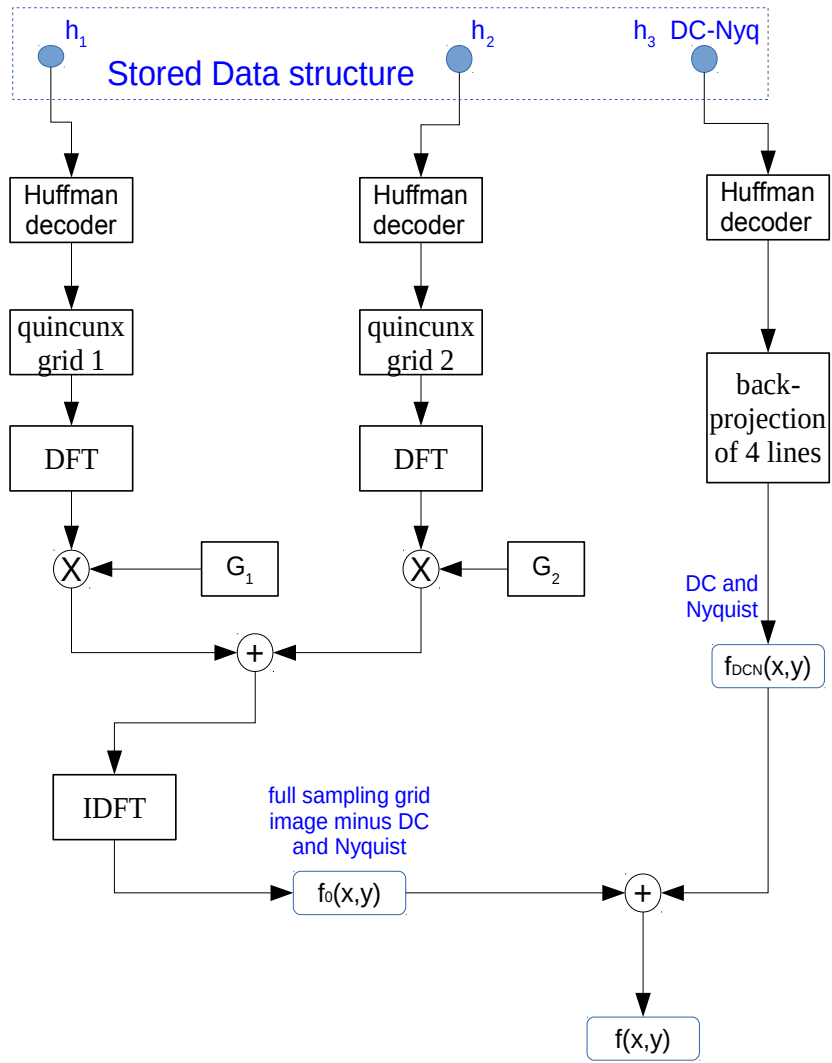

Fig.A1-2
Perfect reconstruction filterbank implementing Papoulis' generalized sampling
for two intertwined partial derivative sample grids and DC-Nyquist side-channel

# Appendix A2: Prior Incarnations and Visualization of The Gradient Mapping

## Prior Incarnations

Although we may be the first to explicitly propose the use of the *gradient mapping* (to use Arnol'd's phrase [49]) ρ(ξ,η) as the joint probability density function in Shannon's entropy equation (page 11, Shannon and Weaver's book [3], many other researchers have investigated rho's theoretical and aesthetic properties. Much of their work has inspired and informed our research. Our work is not complete without a brief review of the main protagonists and their ideas.

Perhaps the earliest recognition of the structural aspects occurred in 19[th] century work on optics. In this field our so-called deldensities are known as caustics. Caustics get their name (Greek for "burning iron") from the curved arc of concentrated light at the focus of a burning glass (or lens). The pre-eminent visual text in this area is Nye's Natural Focusing and Fine Structure of Light [66]. Berry has written at length on many aspects of caustics from geometrical optics [67,68] through wave optics and wave theory [68] into quantum mechanical connections [69]. Upstill investigated perhaps the most ubiquitous source of light caustics: rippling water (and the classic swimming pool effect [70]).

The closely related area of catastrophe theory was originated by Thom [50]. Arnol'd refined the mathematics and used the convenient description *gradient mapping* [71]. Poston and Stewart produced the popular text [72] with wide application to physics, engineering, biology, sociology and ecology; their figure 12.18 on page 257 is a possible incarnation of our proposed image-profiled-lens in section 5.3. If the thickness profile of the irregular lens is defined by an image f(x,y), then the geometric focus at infinity will correspond with a scaled version of equation (3a) and a finite ray-trace spot-diagram will resemble a scaled version of equation (3b).

Astronomers have become interested in the lensing effects of gravitational fields on the appearance of astronomical images. To help interpret the distortions they compute *gravitational lensing magnification map*s [73]. There is a high dimensional computation required to accurately estimate the effect [74]. In many ways the computation resembles the ray-tracing of spot-diagrams in more conventional optical imaging systems.

In 2003 Gluckman proposed a concept he called gradient field distributions for image registration [75]. In Gluckman's figure 1 we immediately recognise our deldensity maps.

Recently there has been some interest shown in quite general caustics known as Argand plane caustics [76,77]. The mapping is generated from the real and imaginary parts of a complex source image. The real part corresponds to the x-partial derivative and the imaginary to the y-partial derivative of our defining equation (3a) and (3b). The idea has recently moved toward *Bloch sphere catastrophes* [78].

In the simplest implementation of our deldensity equation (17) the source image may have perhaps 1 mega-pixels at 8 bit depth. The corresponding delentropy computation in equation (21) only requires a deldensity "image" of 511x511 pixels with an average bin value of about 4. The deldensity "spot-diagram" will have low resolution and will not display clear caustic structures unless the source image is very low bandwidth (compare Fig.2 versus Fig.5).

However, in the continuous limit the fine caustic structure of the deldensity can be expected to not only underpin the mathematics, but also leads to quite splendid graphical objects that adhere to Hardy's mathematical dictum that "beauty is the first test".

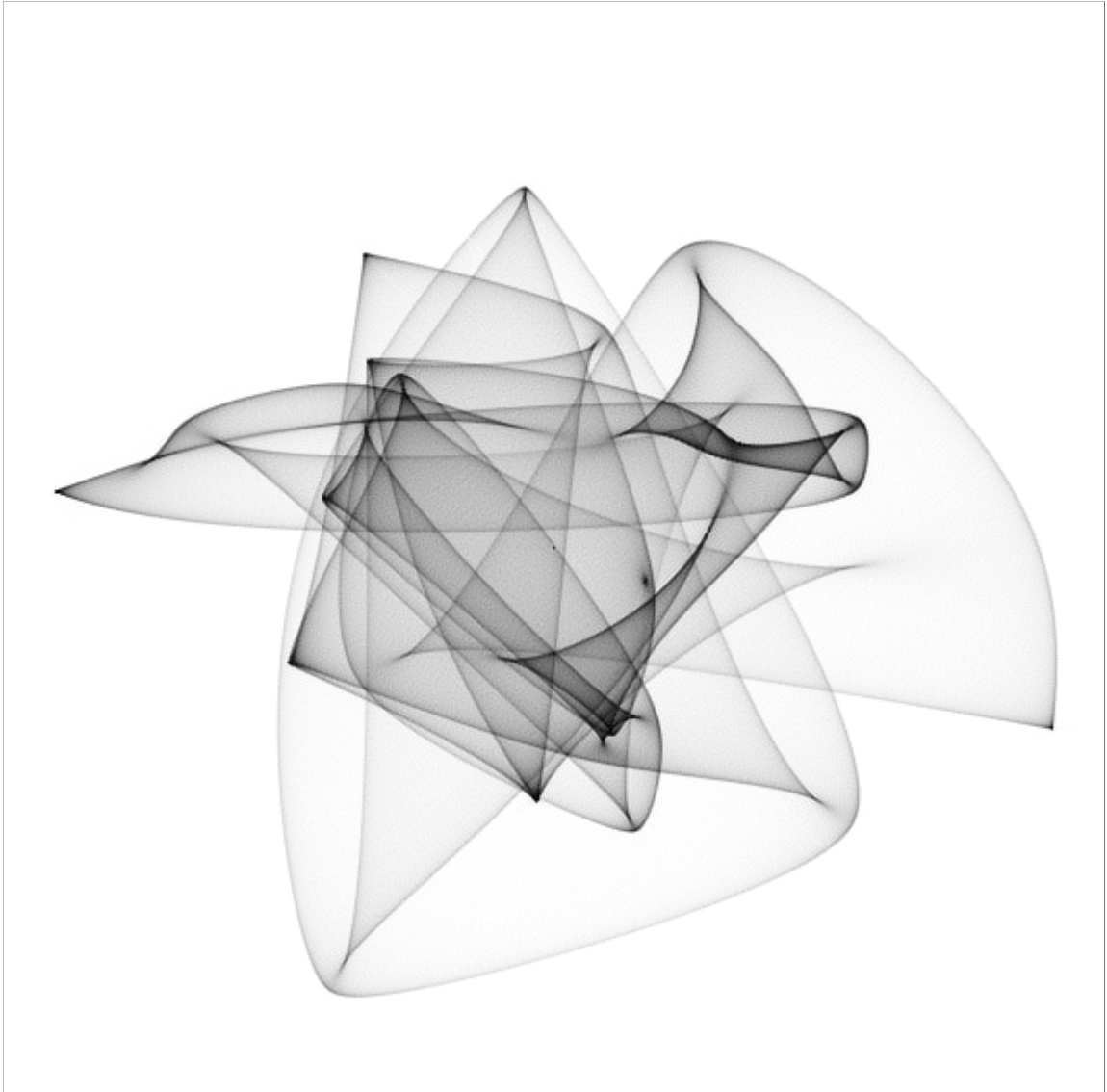

Fig.A2-1
A deldensity map for a low-pass filtered image.  Note the numerous caustics.

## High resolution rendering

Equation (17) implies ray-tracing and histogram binning operations. It defines an algorithm that proceeds pixel-by-pixel and increments a unit value at a bin centered at the discrete coordinates $f_x(m,n)$, $f_y(m,n)$. The most obvious way to obtain high resolution and very fine detail in these ray-trace-like operations is to finely resample the input image and generate floating point gradient estimates. Then the task is to accurately locate the bin either by reducing the bin size or by spreading (splatting) the ray into several adjacent bins based on the offset from bin centre. It turns out that modified sinc splatting can give very fine results in computer graphics [79]. In the related area of optical ray-tracing Gaussian beamlets [80] have been found to give good approximation to wave propagation.

Equation (18) is the continuous domain version of equation (17). It appears impractical for discrete implementation, however we can gain insight by considering the Fourier domain. Starting from equation (18) we can define the 2D Fourier transform:

$$P(\mu,\nu) = \int_{-\infty}^{+\infty}\int_{-\infty}^{+\infty} \rho(\xi,\eta) \exp(-2\pi i[\mu\xi + \nu\eta]) d\xi d\eta$$
$$= \int_{-\infty}^{+\infty}\int_{-\infty}^{+\infty} \left\{ \frac{1}{4AB} \int_{-B}^{+B}\int_{-A}^{+A} \delta(\xi - f_x(x,y), \eta - f_y(x,y)) dxdy \right\} \exp(-2\pi i[\mu\xi + \nu\eta]) d\xi d\eta$$
(A2_1)

The 4D integral conveniently collapses to 2D via the Dirac delta sifting property:

$$P(\mu,\nu) = \frac{1}{4a^2} \int_{-B}^{+B}\int_{-A}^{+A} \exp\left(-2\pi i\left[\mu f_x(x,y) + \nu f_y(x,y)\right]\right) dxdy$$
(A2_2)

$$\rho(\xi,\eta) = \int_{-\infty}^{+\infty}\int_{-\infty}^{+\infty} P(\mu,\nu) \exp(+2\pi i[\mu\xi + \nu\eta]) d\mu d\nu$$
(A2_3)

In words, the 2D Fourier transform of the deledensity, P(u,v), is obtained at every point by a 2D integral of a pure phase function over the image area (A2_2). This suggests a way to get around the discrete binning and achieve a smoother, regularly sampled, near continuous estimate of deldensity.

- Accurately compute partial derivative images on (x,y) grid using Fourier derivative theorem

- Compute equation equation (A2_2) on a regular grid (μ,ν). We can choose fine increments, but note the substantial 2D computation that then results for each point

- Fast Fourier transform P(μ,ν) into ρ(ξ,η) using the FFT. This operation uses a small fraction of resources for of the previous computation.

As far as we know the above computational scheme has not been proposed before. It should compare well with real space versions that use splatting or truncated sinc interpolation. The Fourier approach is approximately equivalent to periodic sinc interpolation in which the sinc is pre-distorted for each and every *splat*. The above Fourier method is ideal in other research fields (i.e. optical ray-tracing, caustics, catastrophe theory and gravity lensing magnification maps) where the density needs to be scrutinised in greater detail.

# Appendix A3: Entropy of vector and phase images in 2D

As we have seen, low redundancy encoding of images can be facilitated by converting a scalar (potential) image into the equivalent vector (field) image and using the two vector components in a joint entropy formalism. The obvious question is then what happens if we consider a vector image directly? Vector and tensor images occur explicitly in many areas of research such as gravitational and magnetic mapping [81,82], flow simulation and medical image processing [83]. Vector imaging also occurs implicitly in areas that use the concept of a phase image such as SAR (synthetic aperture radar) interferometry, optical interferometric imaging [51] and more recently fingerprint compression [52]. To fully define a phase image ψ two components, real and imaginary, must be defined:

$$\exp[i\psi] = \cos\psi + i\sin\psi \tag{A3-1}$$

If we start from a vector image then we can immediately invoke the Fundamental Theorem of Vector Algebra (also known as the Helmholtz-Hodge decomposition to mathematicians) and represent a vector field uniquely as the sum of two distinct parts. (We ignore edge effects in this exposition, but note that these can be included as a harmonic function in a more rigorous formulation.)

In 2D we can write:

$$\mathbf{v}(x,y) = \nabla v_1(x,y) + \nabla \times \{v_1(x,y)\mathbf{k}\} \tag{A3-2}$$
$$= \mathbf{i}\frac{\partial v_1}{\partial x} + \mathbf{j}\frac{\partial v_1}{\partial y} + \left\{\mathbf{i}\frac{\partial v_2}{\partial y} - \mathbf{j}\frac{\partial v_2}{\partial x}\right\}$$

A vector is the sum of the gradient of a scalar potential and the curl of a vector potential. The equivalent for phase images means that a 2D phase can be uniquely decomposed as a curl free part (the scalar potential $v_1$) and a divergence free part (the vector potential $v_2$) respectively [84]. The phases are then isomorphic with potential functions. For example the essence of a human fingerprint image can be captured by a simple cosine fringe pattern [52], based on the observation that a phase modulated image captures all the important first order structure (ridges and minutiae):

$$f(x,y) = 1 + \cos[\psi_1(x,y) + \psi_2(x,y)] \tag{A3-3}$$

Here the potential function $\psi_1$ corresponds with $v_1$ and $\psi_2$ corresponds with $v_2$.

In our initial proposal for 2D delentropy we relied on Papoulis generalised sampling (PGS) to reach usefully low redundancy level (see section 3 and Appendix A1). This is not possible with the more general vector field of equation (A3-2). Essentially PGS uses the strong curl-free constraint on the gradient image field to obtain the 2X sampling advantage. The general 2D vector image has an extra degree of freedom. It can be considered as two independent scalar images simply taking the x and y vector components. However these components are in some sense arbitrary and certainly not invariant to rotation. We can conveniently separate the vector field using the Helmholtz-Hodge decomposition implied by equation (A3-1). In practice we utilise unitary Leray-projection using forward and inverse Riesz transforms to separate the curl-free and divergence-free potentials. In 2D it is even easier to use a complexified vector formalism [51] ideally suited to the many DFT operations required.

Once a vector image has been uniquely decomposed we can apply our new delentropy formalism to each component separately, resulting in a *unique, invariant entropy decomposition for 2D vector fields*.

In the past one very active area of data compression research was the human fingerprint image. An early demonstration of the extreme redundancy of these implicit vector fields was given by the coherent fingerprint formalism [52]. However the data compression was formulated before the delentropy/deldensity concept was known. The Helmholtz-Hodge decomposition separates the phase into curl-free and div-free phase potentials (see images 4 and 5 in [52]). If we apply our deldensity first to the fingerprint continuous (curl-free potential) phase, a phase related to the classic ridge structure, we obtain almost constant gradient magnitude, which translates into low delentropy. The spiral phase (div-free potential) benefits from the del (curl) operator being applied twice (equivalent to a third order isotropic Shannon entropy) resulting in a very sparse map of positive and negative minutiae charges in an ocean of zeros; an extremely low delentropy image. Data compression factors in the hundreds are possible.

In summary it seems that the delentropy formalism developed in the main body of this paper has important (uniquely separable) properties for vector and phase imaging. Rather surprisingly it seems that a third order isotropic Shannon-like delentropy has already been implicated in the extreme compression of fingerprint images.